\documentclass[aip, reprint, jcp]{revtex4-1}

\usepackage{xcolor,hyperref}
\hypersetup{
   colorlinks,
   linkcolor={blue!50!black},
   citecolor={blue!50!black},
   urlcolor={blue!80!black}
}

\usepackage{amsmath}
\usepackage{amssymb}
\usepackage{color}
\usepackage{numprint}
\usepackage{physics}
\usepackage{hyperref}
\usepackage{bm}
\usepackage{graphicx}
\usepackage{placeins}

\npthousandsep{\,}
\DeclareMathAlphabet{\mathitbf}{OML}{cmm}{b}{it}

\newcommand{\xv}{\mathitbf x}

\newcommand{\calBold}[1]{\mbox{\boldmath${\cal #1}$}}
\newcommand{\mathBold}[1]{\mbox{\boldmath$#1$}}
\newcommand{\dbar}{{\,\mathchar'26\mkern-12mu d}}

\begin{document}
\title{Elastic moduli fluctuations predict wave attenuation rates in glasses}

\author{Geert Kapteijns}
\affiliation{Institute for Theoretical Physics, University of Amsterdam, Science Park 904, Amsterdam, Netherlands}
\author{David Richard}
\affiliation{Institute for Theoretical Physics, University of Amsterdam, Science Park 904, Amsterdam, Netherlands}
\author{Eran Bouchbinder}
\affiliation{Chemical and Biological Physics Department, Weizmann Institute of Science, Rehovot 7610001, Israel}
\author{Edan Lerner}
\email{e.lerner@uva.nl}
\affiliation{Institute for Theoretical Physics, University of Amsterdam, Science Park 904, Amsterdam, Netherlands}

\begin{abstract}
The disorder-induced attenuation of elastic waves is central to the universal low-temperature properties of glasses. Recent literature offers conflicting views on both the scaling of the wave attenuation rate $\Gamma(\omega)$ in the low-frequency limit ($\omega\!\to\!0$), and on its dependence on glass history and properties. A theoretical framework --- termed Fluctuating Elasticity Theory (FET) --- predicts low-frequency Rayleigh scattering scaling in $\dbar$ spatial dimensions, $\Gamma(\omega)\!\sim\!\gamma\,\omega^{\dbar+1}$, where $\gamma\!=\!\gamma(V_{\rm c})$ quantifies the coarse-grained spatial fluctuations of elastic moduli, involving a correlation volume $V_{\rm c}$ that remains debated. Here, using extensive computer simulations, we show that $\Gamma(\omega)\!\sim\!\gamma\,\omega^3$ is asymptotically satisfied in two dimensions ($\dbar\!=\!2$) once $\gamma$ is interpreted in terms of ensemble --- rather than spatial --- averages, where $V_{\rm c}$ is replaced by the system size. In so doing, we also establish that the finite-size ensemble-statistics of elastic moduli is anomalous and related to the universal $\omega^4$ density of states of soft quasilocalized modes. These results not only strongly support FET, but also constitute a strict benchmark for the statistics produced by coarse-graining approaches to the spatial distribution of elastic moduli.
\end{abstract}

\maketitle

\section*{Introduction}
It is well-established that the spatio-mechanical disorder featured by structural glasses leads to the attenuation of long-wavelength pure elastic waves, even if nonlinearities and anharmonicities are entirely neglected. \cite{lemaitre_tanaka_2016,Ikeda_scattering_2018,scattering_jcp,wang2019sound,caroli2019fluct,allesio_log_correction_theory_2020} The physics behind this harmonic-regime attenuation is simple: in the presence of glassy structural disorder, pure waves generally do not constitute eigenstates of an amorphous solid's Hamiltonian. Instead, pure waves of frequency $\omega$ project on a subset of `dirty' (disordered) phonons,\cite{phonon_widths} whose spectral width about $\omega$ determines the pure waves' attenuation rate $\Gamma(\omega)$. \cite{lemaitre_tanaka_2016}

Resolving how the wave attenuation rate $\Gamma(\omega)$ in structural glasses depends on wave frequency $\omega$, and on glass formation history, is key to understanding glassy heat transport, which is known to feature universal low-temperature anomalies.\cite{phillips1981amorphous, Zeller_and_Pohl_prb_1971, pohl_review} Despite substantial experimental, \cite{Zeller_and_Pohl_prb_1971,dietsche1979spect,freeman1986thermal,zaitlin1975phonon, monaco2009breakdown, ruta2012acoustic,baldi2013emergence,baldi2014anharmonic, baldi2011elastic, baldi2010sound, foret2002merging, ruffle2006glass, ruffle2003obs, ruffle2008boson}
theoretical,\cite{Schober_prb_1992,Marruzzo2013,eric_boson_peak_emt,allesio_log_correction_theory_2020} and simulational \cite{lemaitre_tanaka_2016,Ikeda_scattering_2018,scattering_jcp,wang2019sound,caroli2019fluct} efforts to shed light on the physics of wave attenuation in structural glasses, many aspects of the phenomenon itself, \cite{lemaitre_tanaka_2016,scattering_jcp,grzegorz_soft_matter_2020,Mossa_2020_prb} and its statistical-mechanical origin, \cite{Schober_prb_1992,Marruzzo2013,eric_boson_peak_emt,Ikeda_scattering_2018,Grzegorz_coarse_graining_soft_matter_2020} remain controversial.

One prominent theoretical framework --- the Fluctuating Elasticity Theory (FET) \cite{footnote4} developed by Schirmacher and coworkers \cite{Schirmacher_2006, Schirmacher_prl_2007, Marruzzo2013} --- predicts that, in the low-frequency/long wavelength limit, the transverse wave attenuation rate obeys~\emph{Rayleigh} scattering scaling,\cite{rayleigh_original} 
\begin{equation}\label{eq:FET}
    \Gamma(\omega)/\omega_0 \,\propto\, \gamma(V_{\rm c})\,(\omega/\omega_0)^{\dbar+1}\,,
\end{equation}
where $\omega_0$ is a characteristic (elastic) frequency scale, and $\dbar$ is the spatial dimension. The dimensionless prefactor $\gamma(V_{\rm c})$ of the Rayleigh scaling --- coined the \emph{disorder parameter} \cite{Schirmacher_2006} --- is defined as
\begin{equation}\label{eq:gamma_def}
    \gamma(V_{\rm c}) \equiv \bigg(\frac{\Delta \mu(V_{\rm c})}{\mu}\bigg)^2 \,\frac{V_{\rm c}}{a_0^\dbar}\,,
\end{equation}
where $\mu$ denotes the macroscopic shear modulus, $\Delta\mu(V_{\rm c})$ denotes the standard deviation of spatial fluctuations of the shear modulus \emph{field} --- coarse-grained on the correlation volume $V_{\rm c}$ ---, and $a_0$ is an interparticle length.\cite{transverse_explanation} 

Equations~(\ref{eq:FET}) and (\ref{eq:gamma_def}) have been recently tested using numerical simulations in two and three dimensions (2D and 3D) in Refs.~\citenum{lemaitre_tanaka_2016,caroli2019fluct,wang2019sound,grzegorz_soft_matter_2020,Grzegorz_coarse_graining_soft_matter_2020}, all of which deemed them either incorrect or incomplete. In Ref.~\citenum{lemaitre_tanaka_2016} it was claimed, based on simulational and some experimental data, that the low-frequency form of the attenuation rate follows $\omega^{\dbar+1}\log(\omega_0/\omega)$ instead of the generic $\omega^{\dbar+1}$ Rayleigh scaling, as a result of long-range spatial correlations in some combinations of first and second order elastic moduli fields. This claim was recently further substantiated by a mean-field theory \cite{allesio_log_correction_theory_2020} that predicts that a logarithmic correction arises whenever long-range correlations in either the elastic constants or internal stresses exist.

Some doubts were, however, raised in Ref.~\citenum{caroli2019fluct} regarding the possibility that correlations in coarse-grained elastic moduli fields give rise to the anomalous, log-corrected scaling, and see also Refs.~\citenum{Ikeda_scattering_2018,scattering_jcp}. In Ref.~\citenum{caroli2019fluct}, it was also shown that a FET framework that neglects disorder-induced non-affine motions underestimates the wave attenuation rate by two orders of magnitude. In Ref.~\citenum{wang2019sound}, some evidence for Rayleigh scaling of $\Gamma(\omega)$ at low frequencies was put forward (see discussion in Ref.~\citenum{scattering_jcp}); in the same work, however, it was also concluded that FET is not quantitatively predictive, based on the apparent failure of Eq.~(\ref{eq:gamma_def}) to account for thermal-annealing-induced variations in $\Gamma(\omega)$. Similar claims were made in Refs.~\citenum{grzegorz_soft_matter_2020,Grzegorz_coarse_graining_soft_matter_2020}.

In this Communication, we provide strong evidence that FET is, in fact, \emph{quantitatively} predictive of long-wavelength wave attenuation rates in structural glasses. Our conclusion is based on the key assumption that coarse-grained local elastic moduli fields do not feature long-range (power-law) correlations, as previously shown using computer simulations in Refs.~\citenum{barrat_moduli_correlations_pre_2009,mizuno2019impact,Grzegorz_coarse_graining_soft_matter_2020} (see, however, claims in Ref.~\citenum{lemaitre_tanaka_2016}). Under this assumption, we replace spatial averages with \emph{ensemble} averages in the definition of $\gamma$ (cf.~Eq.~(\ref{eq:gamma_def})). By doing so, we circumvent the long-standing conundrum of how elastic moduli fields should be defined and coarse-grained, \cite{Goldhirsch2002,barrat_moduli_correlations_pre_2009,barrat_pre_2013_coarse_graining,mizuno2019impact,Grzegorz_coarse_graining_soft_matter_2020} and how their correlation volume is identified. Our results also establish that finite systems follow non-Gaussian elastic moduli statistics, featuring anomalous power-law tails. The latter echo the universal form of the density of states of soft, quasilocalized modes, \cite{soft_potential_model_1991,Gurevich2003,modes_prl_2016,modes_prl_2018,modes_prl_2020} providing an interesting link between micro- and macro-elastic observables. Finally, our results provide a strict benchmark for formulating coarse-graining approaches to elastic moduli fields. 

\section*{Computer glass model}
We employ a generic 50:50 binary mixture 2D glass-forming model, in which pairs $i,j$ of particles at distance $r_{ij}$ from each other interact via a spherically-symmetric, purely repulsive potential $\varphi_{ij}(r_{ij})$ (thus $\varphi'_{ij}\!<\!0$ for all pairs $i,j$), such that the total potential energy reads $U\!=\!\sum_{i<j}\varphi_{ij}$. Details about the model can be found in Ref.~\citenum{cge_paper}. We measure $\Gamma(\omega)$ in the harmonic approximation at zero temperature, as done e.g.~in Refs.~\citenum{lemaitre_tanaka_2016,scattering_jcp}, using the Hessian matrix $\calBold{M}\!\equiv\!\frac{\partial^2U}{\partial\xv\partial\xv}$ of the potential energy $U(\xv)$ that depends on particle coordinates $\xv$. 

In order to explore glassy systems with different degrees of mechanical disorder, we parameterize the Hessian $\calBold{M}(\delta)$ of our glasses by a dimensionless parameter~$\delta\!\in\![0,1]$; the parameterization reads~\cite{eric_boson_peak_emt,inst_note}
\begin{equation}\label{eq:M_of_delta}
    \calBold{M}(\delta) \equiv \calBold{M}'' + (1-\delta)\calBold{M}'\,,
\end{equation}
where
\begin{equation}
    \calBold{M}'' \equiv \sum_{i<j}\varphi_{ij}''\frac{\partial r_{ij}}{\partial\xv}\frac{\partial r_{ij}}{\partial\xv}\ \enspace \mbox{and}\ \enspace \calBold{M}' \equiv \sum_{i<j}\varphi_{ij}'\frac{\partial^2 r_{ij}}{\partial\xv\partial\xv}
\end{equation}
are the stiffness- and internal-force-related terms of the nonparameterized Hessian, respectively. For $\delta\!=\!0$, $\calBold{M}(\delta)$ identifies with the Hessian of the as-cast glasses, while increasing $\delta$ leads to a suppression of the force term $\calBold{M}'$. This procedure has been shown \cite{inst_note,scattering_jcp} to yield (the harmonic approximation of) glassy solids whose micro- and macro-elastic linear-response properties resemble those of glasses created by quenching deeply supercooled liquids to their underlying inherent states.\cite{cge_paper,boring_paper,pinching_pnas} More specifically, increasing $\delta$ mimics deeper supercooling of glasses' ancestral equilibrium configurations, which, in turn, results in the reduction of mechanical inhomogeneities,\cite{boring_paper,pinching_pnas} as is also shown below. 

In order to access the broadest possible range of mechanical disorder/noise, our original, as-cast glasses are quenched from high-temperature liquid states, above the so-called onset temperature. \cite{landscape_dominated_jeppe_2000,onset_reichman_2004} Anticipating a comparison with the FET predictions of Eqs.~(\ref{eq:FET}) and (\ref{eq:gamma_def}), in what follows we express all frequencies and rates in terms of the ($\delta$-dependent) characteristic frequency scale $\omega_0\!\equiv\!c_s/a_0$. Here $c_s$ is the zero-frequency shear wave speed and the interparticle length $a_0$ is given in terms of the number of particles (system's size) $N$ and the system's volume $V$ as  $a_0\!\equiv\!\sqrt{V/N}$. 

\section*{The 2D transverse wave attenuation rate}
\begin{figure}[htpb]
	\centering
	\includegraphics[width=\columnwidth]{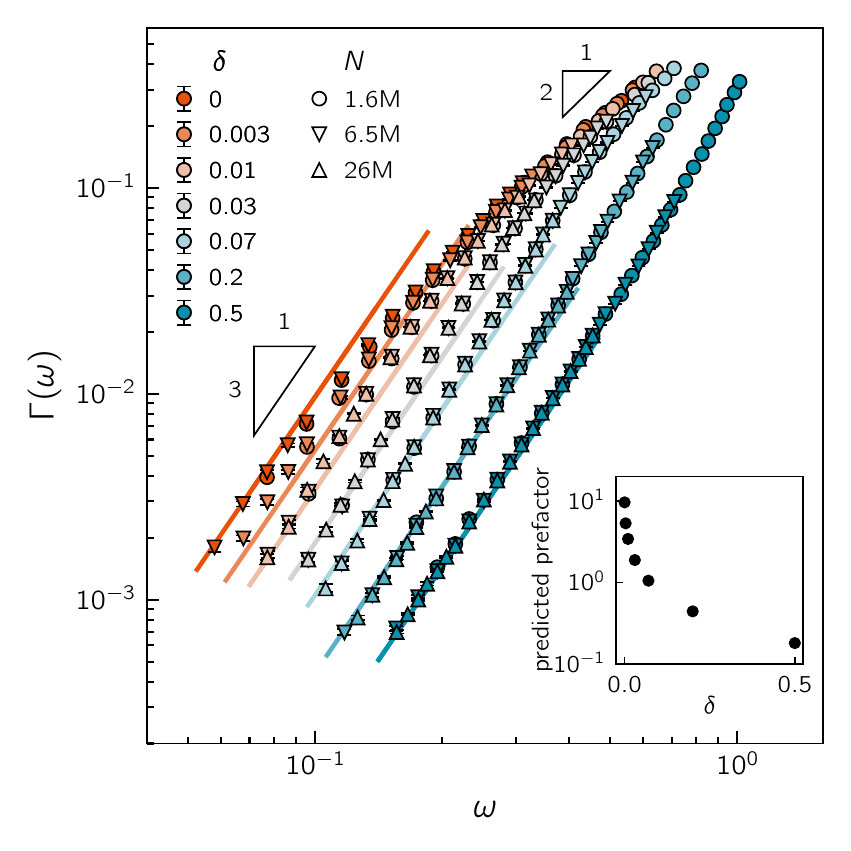}
	\caption{\footnotesize The dimensionless transverse attenuation rate $\Gamma$ vs.~the dimensionless wave frequency $\omega$, for various $\delta$ and $N$ (see legend). The low-frequency Rayleigh-scaling predictions of Eq.~\eqref{eq:FET}, $\sim\!\omega^{\dbar+1}$ (for $\dbar\!=\!2$), are represented by the solid lines, see text for discussion. Also marked is the $\sim\!\omega^2$ high frequency regime. \cite{Schirmacher_prl_2007,eric_boson_peak_emt} Inset: the predicted variation of the Rayleigh-scaling prefactor with $\delta$.}
	\label{fig:fig1}
\end{figure}

We measure $\Gamma(\omega)$ for our glasses under variations of the dimensionless parameter $\delta$ and the system size $N$, while carefully excluding large wavelengths that suffer finite-size effects, as discussed at length in Ref.~\citenum{scattering_jcp}. The results are presented in Fig.~\ref{fig:fig1}; we observe a robust Rayleigh scaling $\Gamma\!\sim\!\omega^3$ at low frequencies for large glasses and all $\delta$ values, and not the log-modified scaling $\sim\!\omega^3\log(\omega_0/\omega)$, even for the as-cast $\delta\!=\!0$ glasses. The continuous lines represent the FET prediction, obtained as explained in what follows. To appreciate the variability of $\Gamma(\omega)$ at low frequencies, we plot in the inset of Fig.~\ref{fig:fig1} the predicted prefactor of the Rayleigh regime vs.~$\delta$, finding a variation of nearly two decades. This large variation surpasses that seen in Ref.~\citenum{wang2019sound} for glasses stabilized by deep supercooling using the SWAP algorithm,\cite{LB_swap_prx} motivating our choice of glass model and $\delta$-procedure.
\begin{figure*}[htpb]
	\centering
	\includegraphics[width=\textwidth]{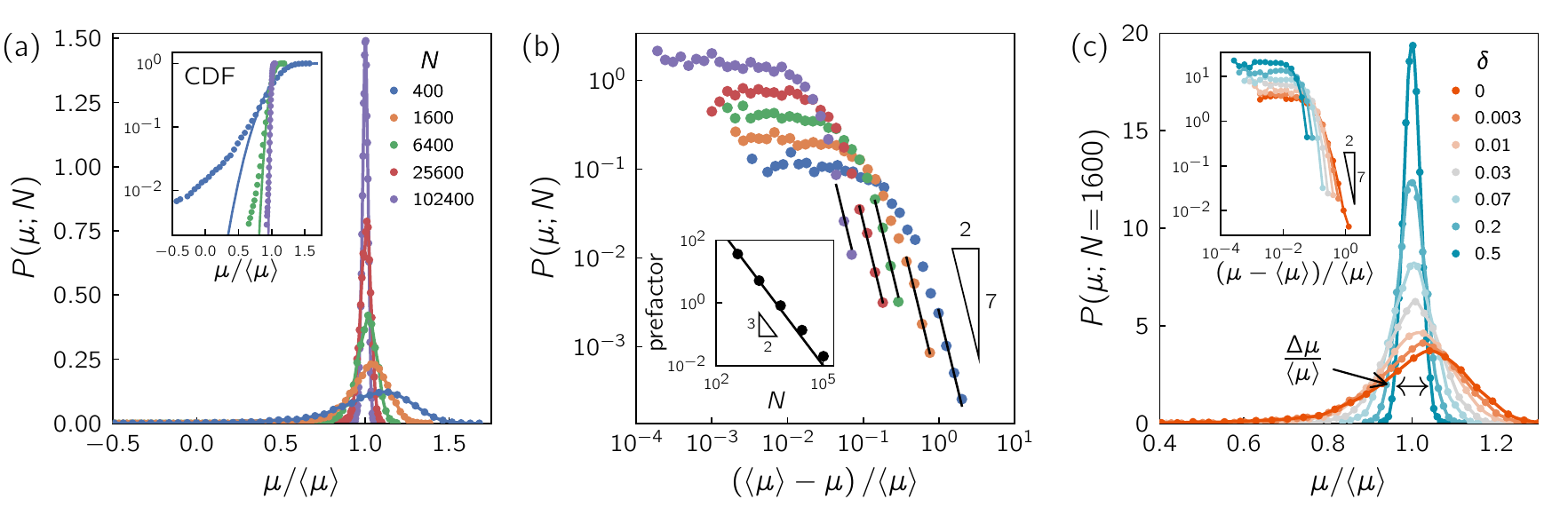}
	\caption{\footnotesize (a) The sample-to-sample shear modulus distribution $P(\mu;N)$ of 10,000 as-cast glasses ($\delta\!=\!0$), for different system sizes $N$ as indicated by the legend. Inset: cumulative distribution function (CDF) of $\mu$, superimposed with Gaussian fits (solid lines), showing that $P(\mu;N)$ features an anomalous tail. (b) Plotting $P(\mu;N)$ vs.~the relative deviation $(\expval{\mu}\!-\!\mu)/\expval{\mu}$ on log-axes reveals the anomalous tails' scaling, which echoes the universal $\omega^4$ distribution of QLMs' frequencies. The inset shows the prefactor of the tails vs.~$N$. (c) $P(\mu;N\!=\!1600$) for various $\delta$s. The inset shows how the anomalous tail gradually disappears for $\delta\!>\!0$.}
	\label{fig:fig2}
\end{figure*}

\section*{Shear modulus fluctuations}
In order to test the FET predictions for $\Gamma(\omega)$, as spelled out in Eqs.~(\ref{eq:FET}) and (\ref{eq:gamma_def}), one would need --- in principle --- to measure coarse-grained moduli \emph{fields} in computer glasses and assess their spatial fluctuations. According to Eq.~(\ref{eq:gamma_def}), one expects that if elastic moduli fields are coarse-grained over a volume $V_{\mbox{\tiny CG}}$ that is sufficiently larger than the moduli's spatial correlation volume $V_{\rm c}$, then $\Delta \mu(V_{\mbox{\tiny CG}})$ scales as $1/\sqrt{V_{\mbox{\tiny CG}}}$. Consequently, $\gamma(V_{\mbox{\tiny CG}})$ is expected to plateau above some correlation volume $V_c$, therefore it can be equivalently assessed by any  coarse-graining volume satisfying $V_{\mbox{\tiny CG}}\!>\!V_{\rm c}$.

The conclusion above suggests that, in the absence of long-range spatial correlations of elastic moduli, one can abandon the coarse-graining program altogether, eliminating any resulting uncontrolled artefacts. Instead, one can consider ensemble --- rather than spatial --- statistics of moduli to assess $\gamma$, where the sample size $N\!\sim\!V$ would play the role of the coarse-graining volume in Eq.~(\ref{eq:gamma_def}). How do sample-to-sample statistics of elastic moduli behave? In Fig.~\ref{fig:fig2}a we show the ensemble-distributions of $\mu$ for as-cast glasses, varying the system size as indicated in the figure legend. The distributions become sharply peaked with increasing $N$, as expected. The fat tails towards $-\infty$, however, are non-Gaussian, as shown by the inset of Fig.~\ref{fig:fig2}a and explained next.

To shed light on the functional form of the leftward tails of the $\mu$ ensemble distributions, we plot in Fig.~\ref{fig:fig2}b the same distributions against $(\langle\mu\rangle\!-\!\mu)/\langle\mu\rangle$ (here $\langle\circ\rangle$ denotes an ensemble average), on logarithmic axes. This representation, together with the inset of Fig.~\ref{fig:fig2}b, suggest that 
\begin{equation}\label{eq:mu_dist}
P(\mu;N)\sim N^{-3/2}\,(\langle\mu\rangle-\mu)^{-7/2}\,,
\end{equation}
for $\mu\!\lesssim\!\langle\mu\rangle\!-\!\Delta\mu$. To understand this anomalous distribution, consider the (athermal) shear modulus, which consists of a difference between two distinct physical contributions: $\mu\!=\!\mu_{\mbox{\tiny BH}}\!- \!\mu_{\mbox{\scriptsize rel}}$. Here $\mu_{\mbox{\tiny BH}}$ is the Born-Huang contribution,\cite{born_book} which is normally-distributed and exists also in ordered systems, and $\mu_{\mbox{\scriptsize rel}}$ is the `relaxation' contribution that is associated with particles' non-affine motions in the presence of disorder.\cite{lutsko,lemaitre2004} The latter takes the form \cite{lutsko,lemaitre2004}
\begin{equation}\label{eq:mu_rel}
    V\mu_{\mbox{\scriptsize rel}} = \frac{\partial^2U}{\partial\epsilon\partial\xv}\cdot\calBold{M}^{-1}\cdot\frac{\partial^2U}{\partial\xv\partial\epsilon}= \sum_\ell \frac{\big(\mathBold{\psi}_\ell\cdot\frac{\partial^2U}{\partial\xv\partial\epsilon}\big)^2}{\omega_\ell^2}\,,
\end{equation}
where $\epsilon$ is a shear strain parameter and $\mathBold{\psi_\ell}$ is the $\ell^{\hbox{\footnotesize th}}$ eigenfunction of $\calBold{M}$ that is associated with the eigenvalue $\omega_\ell^2$ (all masses are set to unity).

The form of $\mu_{\mbox{\scriptsize rel}}$ indicates that low-frequency nonphononic modes can lead to large (negative) contributions to $\mu$ (it is, however, not the case for low-frequency phonons \cite{footnote, exist}). It is now well-established that structural glasses embed a population of soft, quasilocalized modes (QLMs), whose frequencies follow a universal density of states that grows from zero frequency as ${\cal D}(\omega)\!\sim\!\omega^4$ (see Refs.~\citenum{soft_potential_model_1991,Gurevich2003,modes_prl_2016,modes_prl_2018,modes_prl_2020}). Since the deformation couplings $\mathBold{\psi}_\ell\!\cdot\!\frac{\partial^2U}{\partial\xv\partial\epsilon}$ have been shown in Refs.~\citenum{lemaitre_sum_rules_2006,episode_1_2020} to be uncorrelated with the frequencies $\omega_\ell$, and as QLMs' frequencies are largely independent of each other,\cite{footnote2} $\mu_{\mbox{\scriptsize rel}}$ of Eq.~\eqref{eq:mu_rel} can be viewed as the average $\bar{y}$ of ${\cal O}(N)$ independent random variables $y\!\sim\!1/\omega^2$ that are drawn from a power-law distribution $p(y)\!\sim\! {\cal D}\big(\omega(y)\big)|d\omega/dy|\!\sim\! y^{-7/2}$. The heavy-tailed random-walk statistics of $\bar{y}$ has been derived in Ref.~\citenum{lam2011corrections}, precisely mirroring the asymptotic scaling form given by Eq.~(\ref{eq:mu_dist}); namely, a distribution $P(\bar{y};N)\!\sim N^{-3/2}\bar{y}^{-7/2}$ at large $\bar{y}$, quantitatively accounting for the anomalous features of $P(\mu;N)$. 

Despite $P(\bar{y};N)$'s fat, anomalous power-law tail, we have verified that the standard deviation~$\Delta\bar{y}$ of $\bar{y}$ exhibits conventional large-numbers scaling $\sim\!N^{-1/2}$, with no observable finite-size corrections, and thus $(\Delta \bar{y}(N)/\langle\bar{y}\rangle)^2 N\!\sim\!\hbox{const.}$, in analogy with the disorder parameter $\gamma(V_{\mbox{\tiny CG}})$ for $V_{\mbox{\tiny CG}}\!>\!V_{\rm c}$. These results are relevant for as-cast glasses, corresponding to $\delta\!=\!0$. Once $\delta\!>\!0$, a gap $\sim\!\sqrt{\delta}$ is formed in the quasilocalized modes' density of states, as shown in Ref.~\citenum{inst_note}, leading in turn to~the suppression of the power-law tail of $P(\mu;N)$, and to reduced ensemble-fluctuations of $\mu$, as demonstrated in Fig.~\ref{fig:fig2}c.

\section*{Testing the Fluctuating Elasticity Theory}
\begin{figure}[htpb]
	\centering
	\includegraphics[width=\columnwidth]{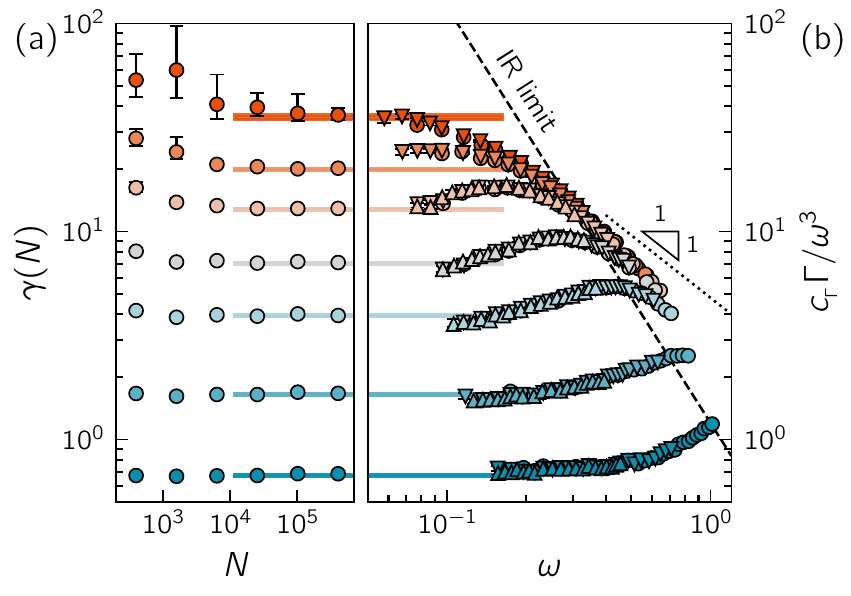}
	\caption{\footnotesize (a) Sample-to-sample disorder parameter $\gamma(N)$ (cf.~Eq.~(\ref{eq:gamma_def2})) vs.~$N$ for different values of $\delta$, represented with the same color code as in Fig.~\ref{fig:fig1}. The colored regions are 95\% confidence intervals of our estimation for $\gamma$ in the large-$N$ limit.\cite{footnote3, bootstrap} (b) Reduced transverse wave attenuation rate $\Gamma/\omega^3$ (scaled by $c_{_\Gamma}\!=\!3.77$) vs.~dimensionless wave frequency $\omega$. Different symbols represent different $N$s as in Fig.~\ref{fig:fig1}. The intersection of the dashed line with $\Gamma/\omega^3$ occurs at the Ioffe-Regel frequency $\omega_{\mbox{\tiny IR}}$ (see text for definition). The dotted line represents the expected high-frequency scaling of $\Gamma/\omega^3$ (see Refs.~\citenum{Schirmacher_prl_2007,eric_boson_peak_emt}).}
	\label{fig:fig3}
\end{figure}

We are now in the position to test the FET predictions, having substituted spatial fluctuations of elastic moduli with their sample-to-sample fluctuations, in the definition of the disorder parameter $\gamma$. In practice, we define the $N$-dependent sample-to-sample disorder parameter
\begin{equation}\label{eq:gamma_def2}
    \gamma(N) \equiv \bigg(\frac{\Delta \mu(N)}{\langle\mu\rangle}\bigg)^2 \, N \,,
\end{equation}
where $\Delta \mu(N)$ denotes the sample-to-sample standard deviation of the shear modulus $\mu$ of glasses of size $N$. Our measurements of $\gamma(N)$ are displayed in Fig.~\ref{fig:fig3}a, for various values of the parameter $\delta$ as indicated by the legend of Fig.~\ref{fig:fig1}, increasing from top to bottom.

We note that, according to the simple random-walk model for the sample-to-sample statistics of $\mu$ proposed above --- which assumes only that $\mu$ is self-averaging and short-range correlated --- we do not expect $\gamma(N)$ to feature an $N$ dependence so long that $N\!>\!V_c/a_0^2$. While we do not attempt to assess $V_c$ here, various previous observations \cite{barrat_moduli_correlations_pre_2009,mizuno2019impact,Grzegorz_coarse_graining_soft_matter_2020} indicate a very safe estimation of the form $V_c/a_0^2\!\lesssim\!10^3$ (in 2D). We nevertheless observe that the lowest-$\delta$'s $\gamma(N)$ features a weak $N$-dependence up to $N\!\sim\!10^5$, which might stem from the tendency of small, highly disordered glassy samples to embed softer excitations than expected,\cite{protocol_prerc,lerner2019finite} leading in turn to larger relative ensemble-fluctuations of $\mu$ at small $N$.

At large $N$, however, $\gamma(N)$ convincingly plateaus for all $\delta$ values, such that the asymptotes provide a prediction for the amplitude of $\Gamma(\omega)$, which is tested next. In Fig.~\ref{fig:fig3}b we plot the measured reduced wave attenuation rate $\Gamma/\omega^3$ (scaled by a numerical proportionality constant $c_{_\Gamma}\!=\!3.77$) against $\omega$. We find a striking agreement between the low-frequency reduced wave scattering rate, and the disorder parameter $\gamma$, as predicted by FET (up to replacing $\gamma(V_c)$ with $\gamma(N\!\to\!\infty)$, cf.~Eqs.~(\ref{eq:gamma_def}) and (\ref{eq:gamma_def2}), and discussions above). That is, 
\begin{equation}
    \Gamma/\omega^3 \sim \gamma(N\to\infty)
\end{equation}
over the entire range of $\delta\!\in\![0,0.5]$, which spans nearly two decades in $\gamma$ and $\Gamma/\omega^3$.

\section*{Summary and discussion}
In this Communication we have shown that the long-wavelength, transverse wave attenuation rate $\Gamma(\omega)$ in 2D glasses follows Rayleigh scaling $\propto\!\omega^3$, with a prefactor proportional to the disorder parameter $\gamma$ (cf.~Eq.~\eqref{eq:gamma_def2}) that quantifies sample-to-sample shear modulus fluctuations. Under a key assumption --- discussed above and further below --- our results support the Fluctuating Elasticity Theory (FET) prediction for long-wavelength attenuation rates, at odds with several recent claims.\cite{lemaitre_tanaka_2016,caroli2019fluct,wang2019sound,grzegorz_soft_matter_2020,Grzegorz_coarse_graining_soft_matter_2020} In order to stringently test the FET predictions, we employed a computer glass in which $\Gamma(\omega)$ at fixed low (dimensionless) frequency can be varied over nearly two decades, by tuning a dimensionless parameter $\delta$ (cf.~Eq.~\eqref{eq:M_of_delta}).

A key assumption we made --- supported by Refs.~\citenum{barrat_moduli_correlations_pre_2009,mizuno2019impact,Grzegorz_coarse_graining_soft_matter_2020} --- is that coarse-grained elastic moduli fields do not feature long-range spatial correlations, and therefore their spatial fluctuations can be equivalently assessed via sample-to-sample fluctuations. Our analysis shows that the sample-to-sample distribution of the shear modulus $\mu$ is non-Gaussian, with an $N$-dependent power-law tail towards negative values, whose exponent echoes the universal $\omega^4$ density of states of soft, quasilocalized modes.\cite{soft_potential_model_1991,Gurevich2003,modes_prl_2016,modes_prl_2018,modes_prl_2020} These anomalies stand at odds with a recent theory of glass elasticity,\cite{Eric_D_field_theory_arXiv_2020} and can be explained via a simple, random-walk model. We stress that any spatial coarse-graining approach to elastic moduli should result in the same anomalous statistics shown here, when the coarse-graining volume is replaced by the system size. 

Several interesting questions emerge from our work. First, the success of FET to predict $\Gamma(\omega)$ over a very broad range of mechanical disorder (as allowed by our 2D glass model) suggests that it should also be predictive in more realistically-formed computer glasses, such as those created with the SWAP algorithm.\cite{LB_swap_prx} This important issue will be addressed in a separate report. Second, to solidify our results, it is crucial to establish whether the proportionality coefficient $c_{_\Gamma}$ between the disorder parameter $\gamma$ and the reduced attenuation rate $\Gamma/\omega^{\dbar+1}$ is universal across models, and to resolve its $\dbar$-dependence.

Finally, $\Gamma(\omega)$ reported in Figs.~\ref{fig:fig1}~and~\ref{fig:fig3} appears to be largely independent of the degree of mechanical disorder in the high-frequency $\Gamma\!\sim\!\omega^2$ regime (i.e.~above the Ioffe-Regel frequency $\omega_{\mbox{\tiny IR}}$ defined via $\pi\Gamma(\omega_{\mbox{\tiny IR}})\!=\!\omega_{\mbox{\tiny IR}}$, see Fig.~\ref{fig:fig3}), consistent with Effective Medium calculations \cite{eric_boson_peak_emt} and with the simulation data of Ref.~\citenum{wang2019sound}. This implies that the reduced rate $\Gamma/\omega^3$ of our maximally-disordered ($\delta\!=\!0$) glasses must \emph{decrease} from the Rayleigh amplitude $\gamma$ to the approximately-$\gamma$-independent $\Gamma/\omega^3\!\sim\!1/\omega$ at $\omega\!\gtrsim\!\omega_{\mbox{\tiny IR}}$, giving rise to an apparent log-corrected scaling $\Gamma/\omega^3\!\sim\!\log(\omega_0/\omega)$ observed first in Ref.~\citenum{lemaitre_tanaka_2016}, and later also in Refs.~\citenum{scattering_jcp,allesio_log_correction_theory_2020}.

Clearly, for more stable glasses featuring substantially smaller disorder parameters $\gamma$, $\Gamma/\omega^3$ will no longer decrease by any appreciable degree from the Rayleigh regime towards the $\omega^2$ regime, ruling out the plausibility of the log-corrected scaling. In addition, we point out that, for the $\delta\!=\!0$ glasses, $\omega_{\mbox{\tiny IR}}$ is merely a factor of $\approx\!3$ higher than the onset of the Rayleigh regime (similar and smaller factors are observed in 3D \cite{scattering_jcp}). All of these issues cast considerable doubt on whether the intermediate frequency regime --- above the Rayleigh regime and below the Ioffe-Regel limit --- can be meaningfully considered as anything other than a crossover between the Rayleigh $\sim\!\omega^{\dbar+1}$ scaling and the disorder-independent, high-frequency $\sim\!\omega^2$ scaling of the wave attenuation rate. 

\section*{Acknowledgements}
We thank Karina Gonz\'alez-L\'opez, Corrado Rainone, Talya Vaknin, Avraham Moriel and Ismani Nieuweboer for their comments on the manuscript. D.~R.~acknowledges support of the Simons Foundation for the ``Cracking the Glass Problem Collaboration" Award No.~348126. E.~B.~acknowledges support from the Minerva Foundation with funding from the Federal German Ministry for Education and Research, the Ben May Center for Chemical Theory and Computation, and the Harold Perlman Family. E.~L.~acknowledges support from the NWO (Vidi grant no.~680-47-554/3259). Parts of this work were carried out on the Dutch national e-infrastructure with the support of SURF Cooperative.

\section*{Data availability}
The data that support the findings of this study are available from the corresponding author upon reasonable request.

\bibliography{scattering_refs}

\begin{thebibliography}{65}
\bibitem [{foo({\natexlab{b}})}]{footnote}%
  \BibitemOpen
  \bibinfo {note} {In the case of phonons
  $\mathBold{\psi}_\ell\!\cdot\!\frac{\partial^2U}{\partial\xv\partial\epsilon}\!\sim\!\omega_\ell$
  \cite{exist}, thus they do not feature large contributions to
  $\mu_{\mbox{\scriptsize rel}}$}\BibitemShut {NoStop}%
\end{thebibliography}

\end{document}